\documentclass[aps,prl,superscriptaddress,reprint,floatfix
]{revtex4-1}

\usepackage{tabularx}
\usepackage{float}
\usepackage{bm}
\usepackage{graphicx}
\usepackage{amsmath}
\usepackage{mathtools}
\usepackage{xcolor}
\usepackage{mathrsfs}
\usepackage[T1]{fontenc}
\usepackage[colorlinks=true,citecolor=blue]{hyperref}
\hypersetup{colorlinks=true,citecolor=blue,linkcolor=blue,urlcolor=blue}

\usepackage{natbib}
\usepackage{comment}

\usepackage{chngcntr}

\begin{document}

\title{Robust polaritons in magnetic monolayers of CrI$_3$}

\author{Yaroslav V. Zhumagulov}
\affiliation{University of Regensburg, Regensburg, 93040, Germany}
\affiliation{HSE University, Moscow, 101000, Russia}

\author{Salvatore Chiavazzo}
\affiliation{Department of Physics and Astronomy, University of Exeter, Stocker Road, Exeter EX4 4QL, United Kingdom}

\author{Ivan A. Shelykh}
\affiliation{Science Institute, University of Iceland, Dunhagi-3, IS-107 Reykjavik, Iceland}
\affiliation{Department of Physics and Engineering, ITMO University, St. Petersburg, 197101, Russia}

\author{Oleksandr Kyriienko}
\affiliation{Department of Physics and Astronomy, University of Exeter, Stocker Road, Exeter EX4 4QL, United Kingdom}

\begin{abstract}
We show that the regime of strong-light matter coupling with remarkable magnetic properties can be realized in systems based on monolayers of chromium triiodide (CrI$_3$). This two-dimensional material combines the presence of ferromagnetic ordering with the possibility of forming strongly-bound excitonic complexes even at room temperature. Using microscopic first-principle calculations we reveal a rich spectrum of optical transitions, corresponding to both Wannier- and Frenkel-type excitons, including those containing electrons with a negative effective mass. We show that excitons of different polarizations efficiently hybridize with a photonic mode of a planar microcavity, and due to the peculiar selection rules polariton modes become well resolved in circular polarizations. The combination of very strong optical oscillator strength of excitons and cavity confinement leads to large values of the Rabi splitting, reaching 35 meV for a single monolayer, and giant Zeeman splitting between polariton modes of up to 20 meV. This makes CrI$_3$ an excellent platform for magnetopolaritonic applications.
\end{abstract}

\maketitle

\textit{Introduction.---}Ensembles of cavity polaritons---composite particles appearing in the regime of strong light-matter coupling---can form quantum fluids of light in optical microcavities. They represent a unique platform for studying collective quantum phenomena at surprisingly high temperatures \cite{CarusottoCuiti2013}. The key characteristic that defines the robustness of polaritonic response is a Rabi splitting $\Omega$ \cite{Khitrova2006}. This parameter is defined by the cavity geometry (e.g. mode volume) and an optical oscillator strength of material excitations participating in polariton formation. For Wannier-Mott excitons in GaAs samples, where the strong coupling was observed for the first time \cite{Weisbuch1992}, the value of $\Omega$ does not exceed few meV. Therefore, while the technology for producing high quality GaAs samples is well developed and III-V polaritons demonstrate excellent properties at cryogenic temperatures, they do not allow for room temperature operation. The search of alternative materials with higher values of $\Omega$ is thus an actual task. Among perspective candidates one should mention wide band semiconductors, such as GaN and ZnO \cite{Zamfirescu2002,Christopoulos2007,Li2013}, organic materials \cite{Lidzey1998,Kena2010,Kena2013,Plumof2014,Betzold2020,Yagafarov2020}, perovskites \cite{Lanty2008,Su2017,Fieramosca2019,Polimeno2020,Su2020}, and carbon nanotubes \cite{Graf2016,Mohl2018,Shahnazaryan2019}. 

An attractive solution is represented by monolayers of transition metal dichalcogenides (TMD) \cite{Splendiani2010}, with typical examples being MoS$_2$, MoSe$_2$, WS$_2$, and WSe$_2$. Their two-dimensional (2D) nature, reduced screening, and relatively high effective masses of carriers lead to a large increase of exciton binding energy and optical oscillator strength \cite{Chernikov2014}. This results in a possibility of boosting the values of $\Omega$ exceeding ten meV \cite{Schwarz2014,LiuMenon2015,Dufferwiel2015,Lundt2017,Schneider2018,Mortensen2020}. Paired with large binding, this makes polaritons robust even at room temperatures \cite{Splendiani2010, Steinleitner2017, Wang2018}. 
\begin{figure}
    \centering
    \includegraphics[width=.5\textwidth]{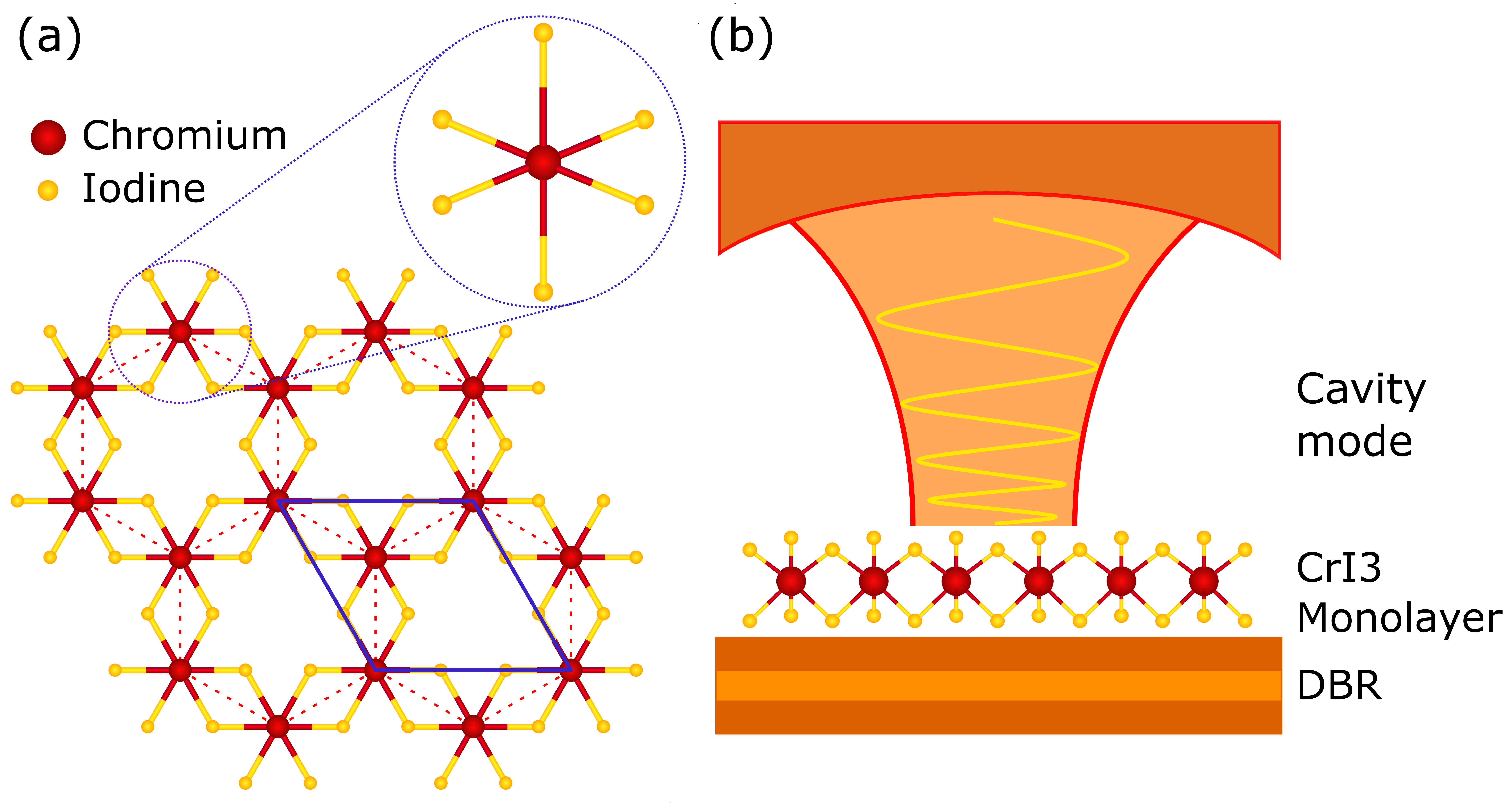}
    \caption{(a) CrI$_3$ monolayer structure. Red and yellow spheres are chromium and iodine atoms. The blue parallelogram and red dashed hexagons highlight unit cells. We enlarge and show a three dimensional visualization of a lattice site, with Cr being in the center, and six iodine atoms surrounding it (shared with other sites). (b) Polaritonic setup where the CrI$_3$ monolayer placed in an optical cavity enters the strong light-matter coupling regime. The cavity is formed by distributed Bragg reflector (DBR) and a fiber-based mirror (top).
    }
    \label{fig:sketch}
\end{figure}
Recently, another class of 2D materials with robust excitonic response, namely chromium trihalides (CrI$_3$ and CrBr$_3$) was discovered.  As compared to typical TMD materials, they have even higher excitonic binding energies and oscillator strengths \cite{Wu2019}. This potentially allows to get Rabi splittings of several tens of meV (see our results). Moreover, an interesting twist is added by the magnetic nature of these and other van der Waals (vdW) materials \cite{Burch2018,Gong2019,Huang2017,Zheng2018,Kashin2020,Cheng2022}. It strongly affects their transport and optical properties, leading to the phenomena of anomalous Hall effect in CrI$_3$ heterostructures \cite{Liu2020}, spin polarized optical conductivity \cite{Kvashnin2020}, giant Kerr rotation \cite{Huang2017}, magnetic circular dichroism \cite{Seyler2018}, onset of 2D magnetoplasmons \cite{Pervishko2020}, and topological excitations \cite{Burch2018}. 

In this paper we analyze the perspectives of the polaritonic applications for chromium trihalides (taking CrI$_3$ as an example, see Fig. \ref{fig:sketch}). We demonstrate that the huge oscillation strength of excitons in these materials allows achieving the Rabi splittings of several tens of meV. Combined with giant values of excitonic Zeeman splitting stemming from the ferromagnetic ordering, this can be of special importance for the observation of polariton topological phases \cite{Nalitov2015,Karzig2015,Klembt2018,Yulin2020}, polaritonic spin Meissner effect \cite{Rubo2006,Walker2011} and valleytronics \cite{Schaibley2016}. Recent results on strong coupling in bulk vdW magnets of nickel phosphorus trisulfide (NiPS$_3$) show that magnetopolaritonics is well within reach of modern experiments \cite{Dirnberger2022}.  

We consider the configuration sketched in Fig. \ref{fig:sketch}(b), where a CrI$_3$ sample is embedded in a planar microcavity. We first address the electron-hole interaction problem and solve a Bethe-Salpeter equation for the spin-resolved interaction Hamiltonian. We reveal a total of four major energy transitions determining the relevant peaks in photoluminescence. Using the Green function approach we then consider the coupling between excitons and cavity photons, and simulate the polaritonic spectra. We demonstrate that due to the peculiar optical selection rules and ferromagnetic properties the polaritonic modes reveal large splitting in circular polarizations.

\textit{Excitons in CrI$_3$ monolayers.---}We start by calculating the properties of the bound states of electron-hole pairs in monolayers of chromium triiodide. The structure of the crystalline lattice  of this material is schematically shown in Fig.~\ref{fig:sketch}(a). 
To concentrate on CrI$_3$ properties, we first consider a pristine monolayer without a substrate and well-separated from other optical components. We note that in practice an exfoliated CrI$_3$ is unstable and requires encapsulation. This can be done by sandwiching the sample between a bulk hexagonal boron nitride (hBN). The presence of hBN adds an extra screening, but does not change qualitatively the excitonic response (we comment on this point when analysing the corresponding spectra). This is similar to the treatment of other TMD monolayers and bilayers \cite{Gerber2019,Mayers2015,Velizhanin2015}.

We calculate the electronic structure of CrI$_3$ monolayers from the first principles using a density functional theory (DFT) that includes both many body interactions and spin-related effects. We perform the calculations using \textsf{GPAW} package \cite{Mortensen_2005,Enkovaara_2010} and utilizing a plane wave basis and LDA functional. The energy cutoff for the basis is set to 600~eV. We use a slab model with a vacuum thickness of 16 {\AA} to avoid interaction between the periodic images. Using the crystal lattice relaxation procedure, we determine the lattice constant for CrI$_3$ being equal to $6.69$ {\AA}. The force convergence criteria is set to 1 meV/{\AA} per atom. Spin-orbit effects are treated at the level of perturbation theory \cite{PhysRevB.94.235106}. To overcome the DFT bandgap problem, we used a scissor operator to achieve the correct value of the bandgap equal to $2.59$~eV, which is determined using the GW approximation \cite{Wu2019}. The exciton spectrum is calculated using the numerical implementation of the Bethe-Salpeter equation, implemented in the \textsf{GPAW} package \cite{PhysRevB.62.4927,PhysRevB.83.245122,PhysRevB.88.245309}, with an additional modification to take into account for spin-orbit interaction in a single-particle basis \cite{PhysRevLett.127.166402}. We calculate the screened potential using the value of the dielectric cutoff equal to 50~eV. A 2D truncated Coulomb potential is applied to achieve fast convergence \cite{PhysRevB.88.245309}. We take into account 200 electronic bands in the calculation of the dielectric function. A set of 16 valence bands and 14 conduction bands on the grid of the Brillouin zone equivalent to 12 by 12 is used as a single-particle exciton basis. In our work, we focus on exciton states with total momentum equal to zero only, as they are most relevant for light-matter coupling in confined photonic structures.
\begin{figure}
    \centering
    \includegraphics[width=.5\textwidth]{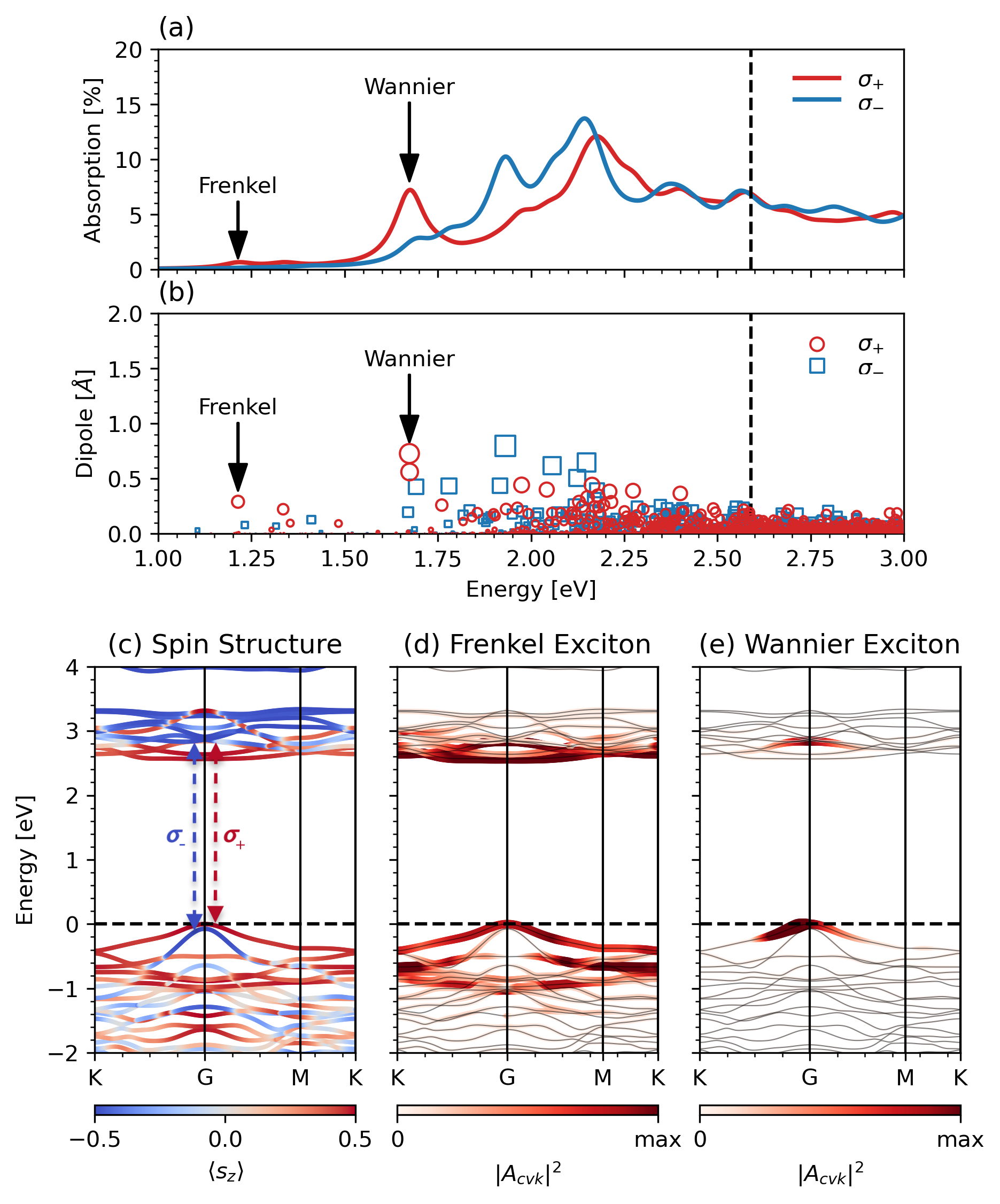}
    \caption{(a) Absorption spectrum of CrI$_3$ resolved in circular polarizations. The red profile corresponds to $\sigma_+$ polarized light and blue profile to $\sigma_-$ polarized light. Lorentzian broadening is taken equal to 80~meV. (b) Scatter plot of excitons dipole moments resolved in circular polarizations. Main peaks are related to the Wannier-type excitons. (c) Spin-resolved band structure of CrI$_3$. Color code denotes spin polarization $\langle s_z \rangle$, shown in the color bar below. Red (blue) dashed two-side arrows show the dominant optical selection rule for $\sigma_+$ ($\sigma_-$) light polarization. (d) Band and momentum single-particle distribution of Frenkel-type excitons, as revealed by their strong delocalization in the momentum space. (e) Band and momentum single-particle distribution of Wannier-type excitons showing the strong localization in momentum space.
    }
    \label{fig:abs}
\end{figure}

The results are presented in Fig.~\ref{fig:abs}. The full absorption profile of the CrI$_3$ sample is presented in Fig.~\ref{fig:abs}(a). The corresponding dipole transition matrix elements are shown in Fig.~\ref{fig:abs}(b). The spectrum is resolved in circular polarizations of light, revealing large non-equivalence between the $\sigma^+$ and $\sigma^-$ components. 
Both $\sigma^+$ and $\sigma^-$ profiles in Fig.~\ref{fig:abs}(a) have strong absorption peaks in the presence of low-lying Frenkel- and Wannier-type excitons. For the $\sigma_+$ polarization, we notice the major transition at $1.7$~eV, while for the $\sigma_-$ polarized radiation the dominant peak is located around $1.9$~eV. 
Fig.~\ref{fig:abs}(b) present the exciton dipole moment distribution in the sample. The optical dipole moments range from below $0.01$~{\AA} to $1.0$~{\AA}, for both $\sigma_+$ and $\sigma_-$ polarizations. 
In Fig.~\ref{fig:abs}(c) we highlight the band spin structure in the momentum space. Conduction and valence band are spin resolved, allowing for the visible spin resolution of excitonic transitions. The main optical selection rules are shown in Fig.~\ref{fig:abs}(c) by dashed two-sided arrows. We note that optical transitions between bands with different spin polarizations are responsible for emitting photons of distinct light polarization. 
\begin{figure}
    \centering
    \includegraphics[width=0.5\textwidth]{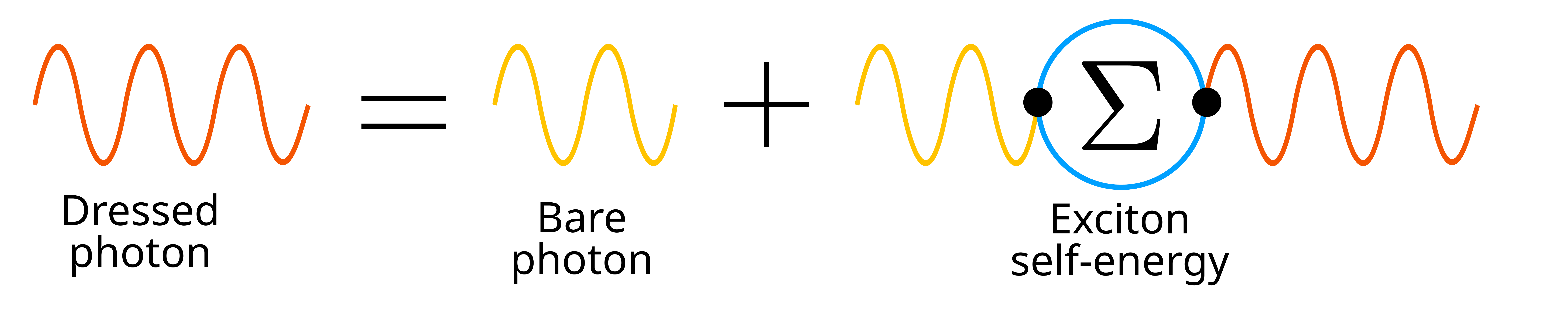}
    \caption{Diagrammatic representation of light-matter coupling described by Eq.~\eqref{eq:greenFunction}. The polaritonic response is described by the dressed photon Green's function, formed by cavity photons that repeatedly interact with CrI$_3$ excitations, described by the excitonic self-energy.}
    \label{fig:selfEnergy}
\end{figure}
Finally, in Fig.~\ref{fig:abs}(d) and Fig.~\ref{fig:abs}(e) we show a typical distribution of the single particle contributions from various bands to Frenkel- and Wannier-type excitons. Frenkel excitons are characterized by a small radius, being of the order of magnitude of the lattice period. In the momentum space they are formed by states originating from multiple bands and wide momentum range. The high density of Frenkel type excitons above $2$~eV [Fig.~\ref{fig:abs}(b)] leads to a blurred spectrum at these energies, thus lowering the resolution of excitonic lines and reducing the effective Rabi splitting. We concentrate on optical properties in the region $\leq 2$~eV, where Wannier type excitons dominate the response, allowing for clear peaks with huge Rabi splitting.

Wannier-type excitons have much bigger Bohr radius, and mostly contain the contributions from two well-defined single bands above and below the gap. The exciton at $1.7$~eV is formed by holes from the top of the upper valence band and electrons from the third conduction band, which have negative effective mass at $\bf{k}=0$ [see Fig.~\ref{fig:abs}(e)]. This is a unique feature of CrI$_3$ monolayers. Even though such band arrangement does not match the most common configuration, for which effective mass in the conduction band is positive, we can readily understand the origin of such a bound state qualitatively within the effective mass description of 2D excitons \cite{HaugKoch}. Considering a two-band model, the Hamiltonian of an electron-hole pair in the center of mass frame $\mathcal H_X$ reads
\begin{equation}
\label{eq:Hx}
    \mathcal H_X = - \frac{\hbar^2}{2 \mu_{cv}} \nabla^2 - V(\vert \vec r \vert),
\end{equation}
with $\vec r$ being the electron-hole relative position, $\mu_{cv}$ the reduced mass, and $V(\vert \vec r \vert)$ the attractive potential. Solutions to Eq.~\eqref{eq:Hx} are bound states when $\mu_{c,v} > 0$. By writing explicitly $\mu_{c,v}$ with negative electron effective mass $m_c=-|m_c|$, $\mu_{cv} = (m^{-1}_v - m^{-1}_c)^{-1}$, we get the condition for getting a bound state as $|m_c| > m_v$. The same constraint was discussed in Refs.~\cite{Lin_2019,Lin_2021,Lin_2021_a}.


\textit{Polaritons in CrI$_3$ monolayers.---}We proceed to study the light-matter coupling in CrI$_3$ and its underlying polarization dependence. We consider a microcavity system with the CrI$_3$ layer embedded between cavity mirrors, as sketched in Fig.~\ref{fig:sketch}(b). 
\begin{figure}
    \centering
    \includegraphics[width=0.5 \textwidth]{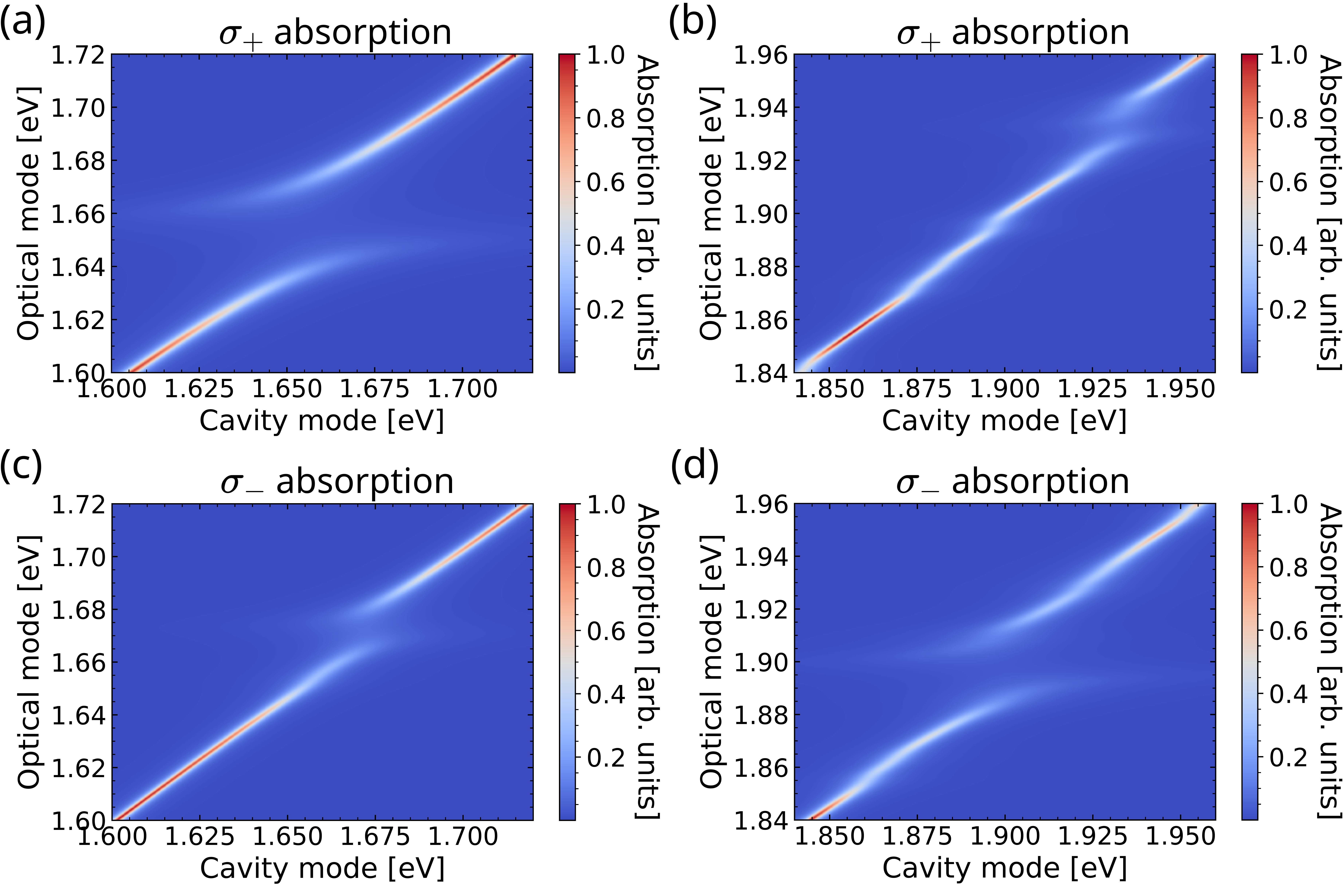}
    \caption{Polarization-resolved CrI$_3$ polariton absorption in two different energy regions. (a, b) $\sigma_+$ absorption profile revealing the presence of Rabi splittings of 32 meV and 14 meV around $1.66$ eV and $1.93$~eV, respectively. For the latter case the splitting is substantially reduced due to the high density of non-degenerate exciton modes. (c, d) $\sigma_-$ absorption profile revealing the presence of Rabi splittings of 16 meV and 34 meV for excitonic transitions at $1.67$~eV and $1.90$~eV, respectively.}
    \label{fig:my_label}
\end{figure}
To study polarization-resolved photoluminescence we assume that excitons are coupled to the single photonic mode which has an antinode in the position where the monolayer is located, and use the Green's function approach. The cavity photon propagator is expressed in the form of $2\times 2$ matrix $\hat{G}^{\mathrm{ph}}(\omega,\textbf{q})$, describing two polarization components \cite{Kyriienko2012,Kyriienko2013,Zhumagulov2022}. Here $\omega$ and $\bf q$ are the photon frequency and momentum, respectively. We account for the dressing of the cavity photons by excitonic quasiparticles in a non-perturbative way, and solve a Dyson-type equation for renormalized photon propagator $\hat{G}^{\mathrm{ph}}(\omega,\textbf{q})$ \cite{Zhumagulov2022,Arnardottir2013} which is graphically shown in Fig.~\ref{fig:selfEnergy}. The renormalized Green's function reads
\begin{equation}\label{eq:greenFunction}
\hat{G}^{\mathrm{ph}}(\omega,\textbf{q})=\frac{\hat{G}^{\mathrm{ph}}_0(\omega,\textbf{q})}{1-\hat{\Sigma}(\omega,\textbf{q})\hat{G}^{\mathrm{ph}}_0(\omega,\bf{q})},
\end{equation}
where $\hat{\Sigma}(\omega,\textbf{q})$ denotes the exciton self-energy obtained from the GW calculations. To get the dispersions of polaritonic modes, we look at the poles of the photon Green's function $\hat{G}^{\mathrm{ph}}$ dressed by material excitations. In the circular polarization basis the Green's function of a bare cavity photon reads
\begin{equation}
\hat{G}^{\mathrm{ph}}_0(\omega,\textbf{q}) = \frac{1}{2}\left[(G^{\mathrm{ph}}_{\mathrm{TE}}+G^{\mathrm{ph}}_{\mathrm{TM}}) \hat \sigma_0 +(G^{\mathrm{ph}}_{\mathrm{TE}}-G^{\mathrm{ph}}_{\mathrm{TM}})\hat{\sigma}_x\right],
\end{equation}
where $\sigma_{0}$ is the identity matrix and the $\sigma_{x}$ is the Pauli matrix. The bare photon Green functions in TE and TM polarizations are
\begin{equation}
G^{\mathrm{ph}}_{\mathrm{TE,TM}}= \frac{2\omega_{\mathrm{TE,TM}}(\textbf{q})}{\omega^2-\omega_{\mathrm{TE,TM}}^2(\textbf{q})+2i\gamma_c\omega_{\mathrm{TE,TM}}(\textbf{q})},
\end{equation}
with photonic dispersion being $\omega_{\mathrm{TE,TM}}(\textbf{q})\approx\omega_0+\hbar^2\mathbf{q}^2/2m_{\mathrm{TE,TM}}$. Here, $\omega_0$ is the resonant frequency of a cavity at normal incidence, $m_{\mathrm{TE,TM}}$ are the effective masses of TE and TM polarized cavity photons, and $\gamma_c$ is the broadening of the photonic linewidth due to a finite transmissivity of the cavity mirrors. 

We approximate the self energy $\hat{\Sigma}(\omega, \mathbf q)$ corresponding to excitonic transitions in both $\sigma^+$  and $\sigma^-$ polarizations as the sum of the Lorentzians corresponding to individual excitons. Accounting for their polarization state, in the matrix form the self energy reads
\begin{equation}
\hat{\Sigma}(\omega,\textbf{q})=\frac{1}{2}\sum_{j}\frac{|\Omega_{j,\mathbf{0}}|^2}{\omega-\omega_j(\textbf{q})+i\gamma_j}\cdot [1+s\hat{\sigma}_x],
\end{equation}
where $s=\pm1$ correspond to two circular polarizations, $\Omega_{j, \mathbf{0}}$ are Rabi energies of the transitions, $\omega_j(\textbf{q})$ are their resonant frequencies, $\gamma_j$ are non-radiative broadenings that depend on a sample quality.

Finally, the Rabi energies for excitonic transitions are calculated as~\cite{HaugKoch, Zhumagulov2022}
\begin{align}
\label{eq:Rabi_def}
    \Omega_{\nu, \mathbf{q}} = -\frac{2i}{\hbar} \sum_{\mathbf{k}} d_{cv}(\mathbf{k}+\mathbf{q}, \mathbf{k})  \sqrt{\frac{\hbar \omega_{\mathrm{c}}(q)}{2 \varepsilon \varepsilon_0 A}} \langle \nu |\mathbf{k}+\mathbf{q}, \mathbf{k}\rangle,
\end{align}
where $d_{cv}(\mathbf{k} + \mathbf{q}, \mathbf{k})$ is a transition dipole matrix element, $\varepsilon$ is a sample permittivity, and $A$ is the sample area.
\begin{figure}
    \centering
    \includegraphics[width=.5 \textwidth]{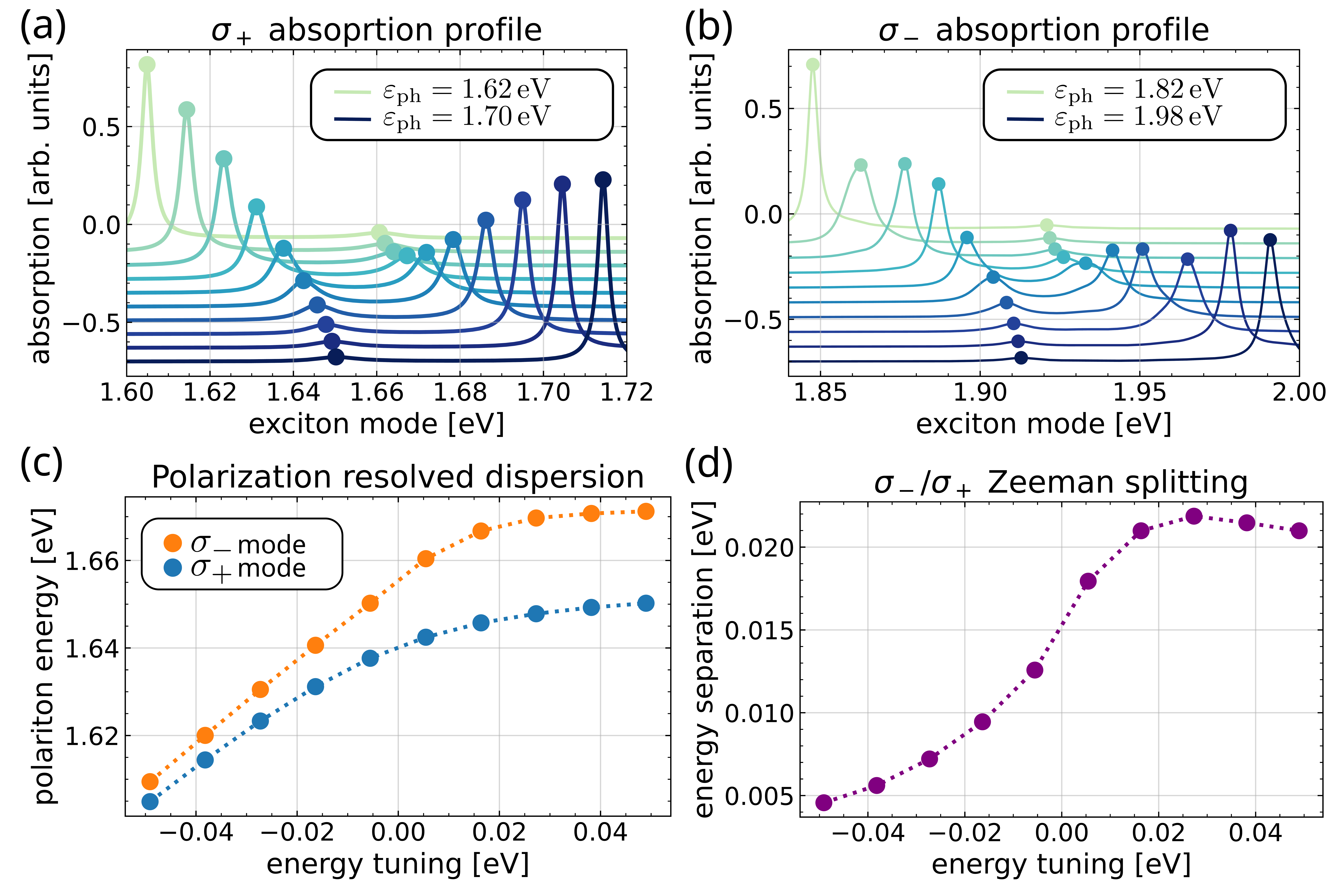}
    \caption{(a, b) Polariton absorption at different photon cavity detunings. (a) $\sigma_+$ absorption with Rabi splitting $\Omega = 32$~meV corresponding to the coupling of cavity photons with excitonic transition at $1.66$~eV. Different curves correspond to different energies of cavity mode, which changes from $1.62$~eV to $1.7$~eV. (b) $\sigma_-$ absorption with Rabi splitting $\Omega = 34$~meV corresponding to the coupling of cavity photons with excitonic transition at $1.93$~eV. Different curves correspond to different energies of cavity mode, which changes from $1.84$~eV up to $1.98$~eV. (c) Lower polariton modes for $\sigma_-$ (orange) and $\sigma_+$ (blue) polarizations, corresponding to the coupling of cavity photons with excitonic transition at $1.66$~eV. Plots are shown as a function of the detuning between exciton and cavity photon energies. (d) Zeeman splitting between $\sigma_-$ and $\sigma_+$ lower polariton modes from the panel (c). Due to the presence of ferromagnetic ordering, the corresponding characteristic value of about $15$~meV at zero detuning.} 
    \label{fig:polaritons}
\end{figure}

In Fig.~\ref{fig:my_label} we show the polarization-resolved absorption spectra in the two energy regions, $1.6$--$1.72$~eV and $1.84$--$1.96$~eV. Due to the ferromagnetic nature of CrI$_3$, the optical response of $\sigma_+$ and $\sigma_-$ polarizations is qualitatively distinct. In the region of $1.6$--$1.72$~eV, opposite circular polarizations show different Rabi splittings, as it can be clearly seen in Fig.~\ref{fig:my_label}(a) and Fig.~\ref{fig:my_label}(c). This leads to the significant polariton Zeeman splitting of the order of 10~meV. In the energy region $1.84$-$1.96$~eV [Figs.~\ref{fig:my_label}(b,d)], the splitting between $\sigma_+$ and $\sigma_-$ lower polaritons becomes even stronger. This is due to the high density of excitonic states in this region and the oscillator strength being redistributed between non-degenerate modes.

We further highlight the two Rabi splittings (major $\sigma_{\pm}$ peaks) in Fig.~\ref{fig:polaritons}(a) and (b), where we show the absorption spectra at different cavity detunings (curves are displaced vertically for clarity) in the two energy regions, where $\sigma_+$ and $\sigma_-$ absorption dominates. The two peaks of each curve correspond to the lower (left peak) and upper (right peak) polariton branches for the leading polarization component. The minimal distance between the two has the value of $\Omega = 32$ meV for $\sigma_+$ and $\Omega = 34$ meV for $\sigma_-$. Lower polariton branches corresponding to two opposite circular polarizations as functions of cavity detunings corresponding to excitonic peak located around 1.66~eV are shown in Fig.~\ref{fig:polaritons}(c) and the corresponding Zeeman splitting, defined as difference of the energies of the two curves is shown in Fig.~\ref{fig:polaritons}(d). We find that it can reach the value of $20$~meV.

\textit{Magnetopolaritons in CrI$_3$.---} The obtained values of the Rabi splitting for CrI$_3$ based systems exceed those characteristic to other TMD monolayers \cite{Chen2020, Kleemann2017}. Moreover, we additionally report the presence of the giant polarization splitting of the polariton modes in the absence of the external magnetic field, which arises due to the ferromagnetic lattice ordering. 
 
Let us consider the region around the resonance located at $1.66$~eV as an example. One can see from Fig.~\ref{fig:my_label}(a, c) that the corresponding excitons are strongly coupled with $\sigma_+$ polarized photons, which pushes the $\sigma_+$ lower polariton (LP$_+$) mode energy down to $1.64$~eV. On the other hand, coupling with $\sigma_-$ photons is much weaker, so that $\sigma_-$ lower polariton (LP$_-$) mode remains above the LP$_+$ [see Fig.~\ref{fig:polaritons}(c)]. The distance between the two modes (polariton Zeeman splitting $\Delta_Z$) becomes of the order $10$--$20$~meV [Fig.~\ref{fig:polaritons}(d)]. This is about two orders of magnitude bigger then the record values reported both in nonmagnetic cavities based on conventional semiconductors \cite{Pietka2015,Walker2011,Rahimi-Iman2011} and TMD monolayers \cite{Lundt2019,Srivastava2015,Dufferwiel2017,Dufferwiel2018}, and overcomes the values characteristic for magnetic semiconductor samples, where in sub-10 T magnetic fields one can reach $\Delta_{\mathrm{Z}} \approx 5$~meV \cite{Mirek2017,Rousset2017,Brunetti2006}. 
 
The possibility of breaking the symmetry between opposite circular polarizations lies at the core of magnetopolaritonics and allows spinoptronic applications, for instance a polariton Berry phase interferometry \cite{Shelykh2009}. Large values of Zeeman splitting are required for the realization of polariton Chern insulators \cite{Nalitov2015,Bardyn2015,Klembt2018}, where formation of topological edge states can be observed if $\Delta_Z$ exceeds the mode broadening \cite{Solnyshkov2021}. 
Another exciting opportunity corresponds to the possibility of braiding the optical and spin excitations in the system \cite{Cenker2021,PhysRevX.8.041028}. Coherent magnon coupling to light has been proposed as a way for microwave-to-optical conversion devices, but corresponding efficiency is usually small due to the weak coupling of magnons to light \cite{Osada2016,Osada2018}, and can be substantially increased in our system, where bright excitons can serve as the mediators \cite{Kudlis2021}.

\textit{Conclusions.---} In conclusion, we developed a microscopic theory for describing exciton polaritons in magnetic monolayers of CrI$_3$ placed in optical microcavities. We demonstrated that the presence of the robust bright excitonic resonances leads to the formation of well resolved polariton modes with values of the Rabi splitting up to $35$~meV, which is comparable with the values characteristic for TMD-based structures. Moreover, the presence of ferromagnetic ordering results in the huge value of the polariton Zeeman splittings of tens of meV, which makes CrI$_3$ monolayers excellent candidates for 2D magnetopolaritonics.

\textit{Acknowledgements.---} Y.\,V.\,Z. is grateful to Deutsche Forschungsgemeinschaft (DFG, German Research Foundation) SPP 2244 (Project No. 443416183) for the financial support. S.\,C. and O.\,K. acknowledge the support from UK EPSRC New Investigator Award under the Agreement No. EP/V00171X/1, and the NATO Science for Peace and Security project NATO.SPS.MYP.G5860. I.\,A.\,S. acknowledges support from Icelandic Research Fund (project ``Hybrid polaritonics'') and Priority 2030 Federal Academic Leadership Program.

\bibliography{Bibliography}

\end{document}